\documentstyle[preprint,aps,epsf]{revtex}
\tighten

\begin{document}

\draft

\title{
Spin dependent momentum distributions 
of proton-deuteron clusters in $\bbox{^3}$He 
from electron scattering on polarized $\bbox{^3}$He:
theoretical predictions
}

\author{
J.~Golak$^{1,2}$,
W.~Gl\"ockle$^1$,
H.~Kamada$^3$,
H.~Wita\l{}a$^2$,
R.~Skibi\'nski$^2$,
A.~Nogga$^4$
}
\address{$^1$Institut f\"ur Theoretische Physik II,
         Ruhr Universit\"at Bochum, D-44780 Bochum, Germany}
\address{$^2$Institute of Physics, Jagiellonian University,
                    PL-30059 Cracow, Poland}
\address{$^3$ Department of Physics, Faculty of Engineering,
   Kyushu Institute of Technology,
   1-1 Sensuicho, Tobata, Kitakyushu 804-8550, Japan}
\address{$^4$ Department of Physics, University of Arizona, Tucson,
              Arizona 85721, USA}
\date{\today}
\maketitle

\begin{abstract}
The process
$\overrightarrow{^3{\rm He}}(e,e'\vec{p})d $
(or $\overrightarrow{^3{\rm He}}(e,e'\vec{d})p $)
is studied theoretically in a Faddeev treatment with the aim to
have access to the spin-dependent momentum distribution 
of ${\vec p} {\vec d} $ clusters in polarized $^3$He.
Final state interactions and meson exchange currents turn out to have a strong influence in
the considered kinematical regime (below the pion threshold). This precludes the direct
access to the momentum distribution except for small deuteron momenta. 
Nevertheless, the results for the longitudinal and transverse response functions 
are interesting as they reflect our present day 
understanding of the reaction mechanism 
and therefore data would be very useful.
\end{abstract}

\narrowtext

\section{Introduction}
\label{secIN}

With the knowledge of solving precisely few-nucleon equations,
 the availability of high precision nucleon-nucleon
(NN) potentials and the  insight into the electromagnetic nucleonic current
 operator it is seducing to ask
very detailed questions about spin dependent momentum distributions
 inside light nuclei and the way to
access them through electron scattering taking final state interactions
 fully into account. Momentum
distributions of polarized $ {\vec d}\, {\vec p} $ 
clusters in spin orientated $^3$He have
 been studied before, see for instance \cite{Fentometer}. We
address here the question whether these distributions are accessible through
the 
$ \overrightarrow{^3{\rm He}} (e,e' {\vec p})d$ or
$ \overrightarrow{^3{\rm He}} (e,e' {\vec d})p$ processes.
Optimal kinematical conditions are that the polarisations of $^3$He and
 of the knocked out proton (deuteron) and the
momenta of the final proton and deuteron are collinear to the photon
 momentum. As we will show 
the longitudinal and transverse response functions 
will lead, up to known factors, directly to the searched for spin dependent momentum
distribution of the ${\vec p} {\vec d}$
clusters in $^3$He. One can also define a proper asymmetry,
which carries corresponding information. 
Of course this can only be true in plane wave impulse
 approximation and for the
absorption of the photon on a single nucleon. Rescattering effects
in the final state as well as
meson exchange currents will disturb the outcome. The strength of that
disturbance again will depend
on the photon momentum Q with the hope that it decreases with increasing Q. 

We formulate the
electromagnetic process in Sec.~II and also display there the searched
for ${\vec p}{\vec d}$ cluster momentum
distributions of $^3$He. Sec.~III shows our results 
for the
$ \overrightarrow{^3{\rm He}} (e,e' {\vec p})d$ and
$ \overrightarrow{^3{\rm He}} (e,e' {\vec d})p$ processes
based on the AV18 NN potential~\cite{AV18}
and precise solutions
of the corresponding Faddeev equations. Since our predictions depend on
the full dynamics in a highly
nontrivial manner, a future experimental verification
will be an important test for
the understanding of few-nucleon dynamics. We end with a brief summary
in Sec.~IV.

\section{Theory}
\label{secTH}

The spin dependent momentum distribution of proton-deuteron
clusters inside the $^3$He nucleus is defined as:
\begin{equation}
{\cal Y} ( M, M_d, m ; {\vec q}_0 ) \ \equiv \
\left\langle \Psi M \left|
| \phi_d M_d  \rangle 
| {\vec q}_0 \, \frac12 m  \rangle  
\langle {\vec q}_0 \,  \frac12 m |
\langle \phi_d M_d |
\right| \Psi M \right\rangle ,
\label{eq1}
\end{equation}
where $ {\vec q}_0 $ is the proton momentum 
(the deuteron momentum is $-{\vec q}_0 $); $m$, $M_d$ and $M$ are
spin magnetic quantum numbers for the proton, deuteron 
and the considered nucleus, respectively. 

We introduce our standard basis in momentum space~\cite{Gloecklebook}
\begin{eqnarray}
\vert p q \alpha \rangle \ \equiv \  \vert p  q ( l s ) j ( \lambda {1 \over 2} ) J
{\cal J} M ( t { 1\over 2} ) T M_{T} \rangle {\rm ,}
\label{eq2}
\end{eqnarray}
where $p$ and $q$ are magnitudes of Jacobi momenta and the set of discrete quantum
numbers $\alpha$ comprises angular momenta, spins and isospins for a three-nucleon (3N)
system. 
Then 
${\cal Y} ( M, M_d, m ; {\vec q}_0 )$
can be evaluated as 

\begin{eqnarray}
{\cal Y} ( M, M_d, m ; {\vec q}_0 ) \ = \
\left|
\sum_{\alpha} \, 
( \delta_{ l 0 } + \delta_{ l 2 } ) \,
\delta_{ s 1 }
\delta_{ j 1 }
\delta_{ t 0 } \, 
C( 1 I \frac12 ; M_d , M - M_d , M )  \right. \cr
\left.
C( \lambda \frac12 I ; M - M_d - m , m, M - M_d ) \
\int_0^\infty d p \, p^2 \,  \phi_l (p) \,
\langle p q_0 \alpha | \Psi \rangle  \
 Y^\star_{ \lambda , M - M_d - m } ( {\hat q}_0 ) \right| ^2 .
\label{eq3}
\end{eqnarray}

In Eq. (3) $ \langle p q_0 \alpha | \Psi \rangle $ are the partial 
wave projected wave function components of $^3$He in momentum space and $  \phi_l (p) $
are the s- and d-wave components of the deuteron.

Further we rewrite  ${\cal Y} ( M, M_d, m ; {\vec q}_0 )$ as
\begin{eqnarray}
{\cal Y} ( M, M_d, m ; {\vec q}_0 ) \, = \,
\left| 
\sum_{\lambda = 0,2 } \, 
Y_{ \lambda , M - M_d - m } ( {\hat q}_0 ) \,
C( 1 I_\lambda \frac12 ; M_d , M - M_d , M ) \right. \cr
\left. C( \lambda \frac12 I_\lambda ; M - M_d - m , m, M - M_d ) \
\sum_{l = 0,2 } \,
\int_0^\infty d p \, p^2 \,  \phi_l (p) \,
\langle p q_0 \alpha_{l \lambda} | \Psi \rangle  \right| ^2 .
\label{eq3.3}
\end{eqnarray}
and define an auxiliary quantity $ H_\lambda (q_0) $ as

\begin{eqnarray}
H_\lambda (q_0) \,\equiv \, 
\sum_{l = 0,2 } \,
\int_0^\infty d p \, p^2 \,  \phi_l (p) \,
\langle p q_0 \alpha_{l \lambda} | \Psi \rangle \ \ , \ \lambda = 0, 2.
\label{eq3.4}
\end{eqnarray}
Note that the set $ \alpha_{l \lambda} $ contributes only for the deuteron 
quantum numbers 
$s=1$, $j=1$ and $t=0$. Further $I_\lambda = \frac12 $ for $\lambda =0$
and $\frac32 $ for $\lambda =2$.
It is clear that using this quantity $H_\lambda (q_0) $
the spin dependent momentum distribution 
$ {\cal Y} ( M, M_d, m ; {\vec q}_0 ) $
can be constructed for any combination of magnetic 
quantum numbers and direction  ${\hat q}_0$.

In this paper all our 
calculations are based on the NN force AV18~\cite{AV18}.
We display $H_\lambda (q_0) $ in Fig.~1.
Note that $\lambda$ is the relative orbital angular momentum 
of the proton with respect to the deuteron inside $^3$He. 
As we see from Fig.~1, 
the $s$-wave ($\lambda$ = 0) dominates the momentum distribution 
${\cal Y}$ for the small relative momenta and has a node around $q_0$ = 400
MeV/c.  Near that value and above the $s$- and $d$-wave contributions 
are comparable. 

In Fig.~2 we show the quantities 
${\cal Y} ( M, M_d, m ; {\vec q}_0 )$
for ${\vec q}_0$ pointing in the direction of the
spin quantisation axis and
the $^3$He nucleus polarised with $M=1/2$. 
The polarisations of the proton and deuteron
are chosen as
$M_d=0, m=1/2$ and $M_d=1, m=-1/2$, respectively. 
We see an interesting shift in the minima from $q_0$= 300 to 500 MeV/c,
if the polarisation of the
proton (deuteron) switches from a parallel (perpendicular) to an antiparallel
(parallel) orientation
in relation to the spin direction of $^3$He. This strong spin dependence
leads to a pronounced spin asymmetry defined as

\begin{eqnarray}
 A \ \equiv \
\frac 
{ 
{\cal Y} ( M= \frac12, M_d= 0, m= \frac12 ; | {\vec q}_0 | {\hat z} ) - 
{\cal Y} ( M= \frac12, M_d= 1, m= -\frac12 ; | {\vec q}_0 | {\hat z} )
} 
{ 
{\cal Y} ( M= \frac12, M_d= 0, m= \frac12 ; | {\vec q}_0 | {\hat z} ) + 
{\cal Y} ( M= \frac12, M_d= 1, m= -\frac12 ; | {\vec q}_0 | {\hat z} )
} .
\label{eq3.5}
\end{eqnarray}
and shown in Fig.~3.

Next we ask the question, how this quantity can be accessed experimentally.
The cross section for the process
$ e + ^3{\rm He} \rightarrow e' + p + d$ has the form~\cite{Donnelly}

\begin{eqnarray}
\sigma \ = \ 
\sigma_{\rm Mott} \, 
\left\{ \,
\left( v_L W_L + v_T W_T + v_{TT} W_{TT} + v_{TL} W_{TL} \right) \,
\ + \  h \, \left( v_{T'} W_{T'} + v_{TL'} W_{TL'} \right) \,
\right\} \,
\rho ,
\label{eq4}
\end{eqnarray}
where $\sigma_{\rm Mott}$, $v_i$ and $\rho$ are analytically given kinematical factors,
and $h$ is the helicity of the incoming electron.
The response functions $W_i$, 
which contain the whole dynamical information, are constructed from 
the current matrix elements taken between the initial bound state $| \Psi M \rangle $
and the final scattering state $| \Psi_{pd}^{(-)} M_d m \rangle $~\cite{we-electron}. 
They are given as 

\[
W_L =  
\left| \langle \Psi_{pd}^{(-)} M_d m | j_0 ( {\vec Q} ) | \Psi M \rangle \right| ^2  
\ \equiv \ \left| N_0 \right| ^2 ,
\]

\[
W_T =  
\left| \langle \Psi_{pd}^{(-)} M_d m | j_{+1} ( {\vec Q} ) | \Psi M \rangle \right| ^2  \ + \
\left| \langle \Psi_{pd}^{(-)} M_d m | j_{-1} ( {\vec Q} ) | \Psi M \rangle \right| ^2 
\ \equiv \ \left| N_{+1} \right| ^2 \ + \ \left| N_{-1} \right| ^2 ,
\]

\[
W_{TT} =  2 \, \Re ( N_{+1} (N_{-1})^\star ) ,
\]

\[
W_{TL} = -2 \, \Re ( N_0 ( N_{+1} - N_{-1} )^\star ) ,
\]

\[
W_{T'} =  \left| N_{+1} \right| ^2 \ - \ \left| N_{-1} \right| ^2 ,
\]

\begin{eqnarray}
W_{TL'} = -2 \, \Re ( N_0 ( N_{+1} + N_{-1} )^\star )
\label{eq5}
\end{eqnarray}
Note that $W_{T'}$ and $W_{TL'}$ contribute only in the case
when the initial electron is polarized.
This is our standard notation $N$ of the nuclear matrix
 element, where
the indices  $0$ and $\pm 1$ stand for the zeroth component and
 the transverse spherical components of the current.
The general 3N current operator contains the single nucleon 
contributions as well as two- and three-nucleon exchange terms

\begin{eqnarray}
j_\mu ( {\vec Q} ) = j_\mu ( {\vec Q} ; 1 ) + j_\mu ( {\vec Q} ; 2 ) + j_\mu ( {\vec Q} ; 3 ) .
\label{eq6}
\end{eqnarray}

In the nonrelativistic limit, which we use, 
 the three contributing pieces of the single 
nucleon current operator (the charge density, 
the convection and the spin current) 
can be written in the 3N momentum space as 

\begin{eqnarray}
j_0 ( {\vec Q} ; 1 ) = 
\int d {\vec p} \int d {\vec q} \  | \, {\vec p} \, {\vec q} \, \rangle \
\hat{\Pi} ( Q ) \  \langle \, {\vec p} \, {\vec q} - \frac23 {\vec Q} \, | ,
\label{eq7}
\end{eqnarray}
\begin{eqnarray}
j_\tau ( {\vec Q} ; 1 ; {\rm conv} ) = 
\int d {\vec p} \int d {\vec q} \  | \, {\vec p} \, {\vec q} \, \rangle \ \frac{q_\tau}{m_N} \
\hat{\Pi} ( Q ) \  \langle \, {\vec p} \, {\vec q} - \frac23 {\vec Q} \, | ,
\label{eq8}
\end{eqnarray}
\begin{eqnarray}
j_\tau ( {\vec Q} ; 1 ; {\rm spin} ) = 
\int d {\vec p} \int d {\vec q} \  | \, {\vec p} \, {\vec q} \, \rangle \ 
\frac{Q \tau \sigma_\tau}{2 m_N} \
\hat{\Pi}_M ( Q ) \  \langle \, {\vec p} \, {\vec q} - \frac23 {\vec Q} \, | ,
\label{eq9}
\end{eqnarray}
where $m_N$ is the nucleon mass and $ \hat{\Pi} ( Q ) $ and $\hat{\Pi}_M ( Q )$ 
are sums of isospin projection operators for the neutron and proton 
joined by the electric ($G_E$) and magnetic ($G_M$)
nucleon form factors, respectively (see~\cite{we-electron}). 
We assumed that $ {\vec Q} \parallel {\hat z} $.

Let us now decompose the scattering state $| \Psi_{pd}^{(-)} M_d m \rangle $
in the following way

\begin{eqnarray}
| \Psi_{pd}^{(-)} M_d m \rangle  \equiv 
| \phi_{d} M_d {\vec q}_f m \rangle  + | \Psi_{pd}^{\rm rest} M_d m \rangle  .
\label{eq10}
\end{eqnarray}
The first term is just a product of the deuteron wave
function $ | \phi_{d} M_d \rangle $ 
and a relative momentum eigenstate of the spectator nucleon $ | {\vec q}_f m \rangle $.
The other term accounts for the proper antisymmetrization 
of the final state and all rescattering contributions.

If the many-nucleon contributions to the 3N current
($ j_\mu ( {\vec Q} ; 2 ) $ and $ j_\mu ( {\vec Q} ; 3 ) $) 
and  $ | \Psi_{pd}^{\rm rest} M_d m \rangle $
can be neglected (PWIA assumption), 
then the current matrix elements take the following form 

\begin{eqnarray}
N_0^{\rm PWIA} ( M, M_d, m ) \ = \ G_E ( Q ) \, 
\sum_{\alpha} \,
( \delta_{ l 0 } + \delta_{ l 2 } ) \,
\delta_{ s 1 }
\delta_{ j 1 }
\delta_{ t 0 } \
C( 1 I \frac12 ; M_d , M - M_d , M ) \cr
C( \lambda \frac12 I ; M - M_d - m , m, M - M_d ) \,
Y_{ \lambda , M - M_d - m } ( \widehat{{\vec q}_f  - \frac23 {\vec Q}} ) \cr
\int_0^\infty d p \, p^2 \,
\left\langle p \, | {\vec q}_f  - \frac23 {\vec Q} | \, \alpha \left| \right. 
 \Psi \right\rangle 
\, \phi_l (p)
\label{eq11}
\end{eqnarray}

\begin{eqnarray}
N_\tau^{\rm conv \ PWIA} ( M, M_d, m ) \ = \ 
\sqrt{\frac{4 \pi}{3}} \, \frac{q_f}{m_N} \,  
Y_{ 1 \tau } ( {\hat q}_f ) \, 
N_0^{\rm PWIA} ( M, M_d, m ) 
\label{eq12}
\end{eqnarray}

\begin{eqnarray}
N_\tau^{\rm spin \ PWIA} ( M, M_d, m ) \ = \ 
\frac{\sqrt{3}}{2} \, \tau \, \frac{Q}{m_N} \, G_M ( Q ) \, 
C( \frac12 1 \frac12 ; m - \tau , \tau , m ) \cr
\sum_{\alpha} \,
( \delta_{ l 0 } + \delta_{ l 2 } ) \,
\delta_{ s 1 }
\delta_{ j 1 }
\delta_{ t 0 } \
C( 1 I \frac12 ; M_d , M - M_d , M ) \cr
C( \lambda \frac12 I ; M - M_d - m + \tau , m - \tau , M - M_d ) \,
Y_{ \lambda , M - M_d - m + \tau } ( \widehat{{\vec q}_f  - \frac23 {\vec Q}} ) \cr
\int_0^\infty d p \, p^2 \,
\left\langle p \, | {\vec q}_f  - \frac23 {\vec Q} | \, \alpha \left| \right.
 \Psi \right\rangle
\, \phi_l (p)
\label{eq13}
\end{eqnarray}
In the laboratory frame 
$ {\vec p}_N + {\vec p}_d  = {\vec Q} $
and by definition of the Jacobi momentum
$ {\vec q}_f = \frac23 {\vec p}_N - \frac13 {\vec p}_d $,
thus 
$ {\vec q}_f - \frac23 {\vec Q} = -{\vec p}_d $.
The second argument of the $^3$He wave function component is therefore
just the deuteron lab momentum.
For the parallel kinematics ($ {\vec Q} \parallel {\vec p}_N \parallel {\vec p}_d $)
the matrix element $N_\tau^{\rm conv \ PWIA} $ is zero.

In this particular situation, and
for the initial target spin parallel to ${\vec Q} $ ($M = \frac12$)
only few combinations of the magnetic quantum numbers contribute
to the nuclear matrix elements
$N_0^{\rm PWIA}$ and 
$N_{\pm 1}^{\rm spin \ PWIA}$. 
Because of the choice of the parallel kinematics
and the property of the spherical harmonics  
these are 
$M = \frac12 , M_d = 0 , m= \frac12 $ 
and $M = \frac12 , M_d = 1 , m= -\frac12 $ in $N_0^{\rm PWIA}$,
$M = \frac12 , M_d = 0 , m= -\frac12 $ 
and $M = \frac12 , M_d = -1 , m= \frac12 $ in $N_{-1}^{\rm spin \ PWIA}$
and
$M = \frac12 , M_d = 1 , m= \frac12 $ in $N_{+1}^{\rm spin \ PWIA}$.

Furthermore, if we compare the expressions given in Eqs. (14) and (16) to the
one in Eq. (3) we find that the
searched for spin dependent momentum distributions ${\cal Y}$ 
of $^3$He are connected to $N_i^{\rm PWIA}$ by
\begin{eqnarray}
{\cal Y} ( M = \frac12, M_d = 0, m = \frac12 ; | {\vec p}_d | {\hat z} ) \ = \cr
\frac1{ (G_E)^2 } \left| N_0^{\rm PWIA} ( M = \frac12, M_d = 0, m = \frac12 ) \right|^2  \ = \cr
\frac{2 m_N^2}{Q^2 (G_M)^2 } \left| N_{-1}^{\rm spin \ PWIA} ( M = \frac12, M_d = 0, m = -\frac12 ) \right|^2 
\label{eq14}
\end{eqnarray}
and by
\begin{eqnarray}
{\cal Y} ( M = \frac12, M_d = 1, m = -\frac12 ; | {\vec p}_d | {\hat z} ) \ = \cr
\frac1{ (G_E)^2 } \left| N_0^{\rm PWIA} ( M = \frac12, M_d = 1, m = -\frac12 ) \right|^2  \ = \cr
\frac{2 m_N^2}{Q^2 (G_M)^2 } \left| N_{+1}^{\rm spin \ PWIA} ( M = \frac12, M_d = 1, m = \frac12 ) \right|^2 .
\label{eq15}
\end{eqnarray}
In the case of parallel kinematics $W_{TT}$, $W_{TL}$ and $W_{TL'}$ vanish.
This follows from the fact that the conditions on the magnetic quantum
numbers, $M$, $M_d$ and $m$, given in products of $N_0$, $N_{+1}$
and $N_{-1}$ cannot be simultaneously fulfilled.
For an experiment with unpolarized electrons, the cross section (7)
contains then only the longitudinal  ($W_L$) and transverse ($W_T$)
response functions:
\begin{eqnarray}
\sigma \ = \
\sigma_{\rm Mott} \,
\left( v_L W_L + v_T W_T \right) \,
\rho ,
\label{eq45}
\end{eqnarray}
Thus the standard ``L-T'' separation is required in order to access
individually $W_L$ and $W_T$.

Another possibility is offered by an experiment with a polarized electron
beam. In this case no further separation of response functions is required,
since

\begin{eqnarray}
\frac12 ( \sigma(h=+1) - \sigma(h=-1) ) \frac1{v_{T'}\, \rho} \ = \
{\vert N_{+1} \vert}^2 - {\vert N_{-1} \vert}^2 .
\label{eq46}
\end{eqnarray}

Therefore under these extreme simplifying assumptions the response
functions $W_L$, $W_T$ and $W_{T'}$ carry
directly the searched for information. Note that in case of $W_T$ ($W_{T'}$)
only one of the two parts gives a nonzero contribution.

The full dynamics adds antisymmetrization in the final state.
(Note our single nucleon current
operator as given in Eqs.~(10)-(12) acts only on one particle. Antisymmetrization
in the final state is
equivalent to the action of the current on all three particles). Then of
course rescattering to all
orders in the NN $t$-operator has to be included. On top one should add
at least two-body currents. We
have described how to do that before at several places~\cite{we-electron}. 
Here we only remark that we employ standard $\pi$- and $\rho$-like 
exchange currents 
related to the NN force AV18, which we use throughout the paper, 
and that adequate Faddeev
equations for $^3$He and for the treatment of FSI have been solved precisely.

\section{Results}
\label{secRes}

Since we work strictly nonrelativistically we want to keep the 3N c.m. energy,
$E_{3N}^{c.m.}$, below the pion threshold.
But in that regime we would like to study many kinematical configurations and 
also include higher three-momenta $Q$ of the photon.
We display in Table~I the kinematical conditions, for
which our studies have been
carried through. In parallel kinematics one can distinguish three cases
for the momentum orientations of the final proton and deuteron,
which we denote by $C_1$, $C_2$ and $C_3$,
and which are depicted in Fig.~4. Thus for $C_2$ the final momenta
of proton and deuteron are parallel to ${\vec Q}$, 
whereas in $C_1$ and $C_3$ only one of them lies in the direction of ${\vec Q}$, 
the other is opposite. Table~I shows for an (arbitrarily selected) initial electron energy
of 1.2 GeV various relevant variables: the electron scattering angle,
the proton and deuteron momenta $p_N$ and $p_d$,
the photon energy $\omega$,
the three momentum of the photon  $Q$ 
and finally the 3N c.m. energy $E_{3N}^{c.m.}$.
The additional
label distinguishes the three cases $C_1$ to $C_3$. 
We see that for each fixed $p_d$ value we cover a certain range 
of $Q$-values. 
The three $C_1$ configurations with $E_{3N}^{c.m.} > $ 140 MeV
are above the pion threshold and have to be taken with caution.
We evaluated
all the cases of Table~I but do not show all in case the results 
are similar.
Fig.~5 displays
$W_L / (G_E)^2$ for $M_d= 0$, $m= \frac12$ 
and $M_d= 1$, $m=-\frac12$ against the available $Q$-values
according to Table~I. 
According to Eqs.~(17) and~(18) in PWIA 
$W_L / (G_E)^2$ is just the searched for ${\cal Y}$
and thus trivially independent of $Q$. 
Symmetrising the final state but still neglecting rescattering is called 
PWIAS, while predictions including additionally FSI are denoted by Full.
We see a change of patterns in going from $p_d$= 100 to 200 and from 400 to 500 MeV/c. 
As seen  from Table~I this is related to the different 
motions of the final proton and deuteron, in other words one
switches from the configuration $C_1$ to $C_2$ and then to $C_3$. Symmetrisation
(PWIAS) has little effect at $p_d$= 10 (not shown) and 100 MeV/c but has a big one 
for the smaller $Q$ values in case 
of $p_d$= 200--400 MeV/c
and  for all $Q$-values in case of $p_d$= 500--600 MeV/c. 
Rescattering plays mostly
a strong role. In case of $C_1$ ($p_d$= 10 and 100 MeV/c) its effects are relatively
small and diminish nicely with increasing Q. In case of
$C_2$ ($p_d$= 200--400 MeV/c) its role is dramatic for $p_d$= 300 and 400 MeV/c,
which has to be expected
since the proton and the deuteron travel 
together with a low relative energy $E_{3N}^{c.m.}$. 
In case of $C_3$ 
the two particles travel again opposite
to each other as for $C_1$ and $E_{3N}^{c.m.}$ decreases with increasing $Q$. 
In this case the by far dominant contribution to the
very strong deviation from PWIA comes from antisymmetrization in the final
state and FSI leads to a relatively mild modification in case of $m= \frac12$ but
a significantly larger one for $m= -\frac12$.
Thus we see quite
different outcomes depending on the cases and these theoretical predictions
would be very interesting to be compared to data.

In case of $W_T$ and $W_{T'}$ the spin operator appears in the current 
and moreover one can see the
effects of the $\pi$- and $\rho$-like MEC's. 
Nevertheless the situation 
for
$\frac{2 m_N^2}{Q^2 (G_M)^2 } \, W_T$
shown in Fig.~6
is roughly spoken similar to the one for 
$W_L / (G_E)^2$.
(We regard only $W_T$ but of course $W_{T'}$ carries the same information.)
Additionally one observes the effects of MEC's, which are pronounced
for $p_d$= 300 and 400 MeV/c.

In view of all that, can we identify kinematic regions to pin down the searched
for momentum distributions using $W_L$ or $W_T$ ?
We choose the cases of the closest
approach of PWIA and Full calculations (with MEC's in case of $W_T$)
for the different $p_d$ values.
They are displayed in Figs.~7 and~8 
together with the spin dependent momentum distributions ${\cal Y} $
from Fig.~2. 
In case of $m= 1/2$ the values of closest approach extracted from $W_L$ and $W_T$ differ for
the larger $q_0$ values, where they also do not reach ${\cal Y}$. 
Only to the left of the zero of ${\cal Y}$
they agree with each other and with  ${\cal Y}$. 
For $m= -1/2$ the predictions for $W_L$ and $W_T$ agree with
each other but do not show the strong dip of  ${\cal Y}$. 
For the smaller $q_0$-values they agree with  ${\cal Y}$.
As a consequence of these results  the asymmetry $A$ formed out of
those values of closest approach cannot follow the asymmetry 
formed out of the  ${\cal Y}$'s. Only the
values extracted for $W_L$ show a mild similarity with the searched for $A$,
as  shown in Fig.~9.

\section{Summary}
\label{secSum}

Based on the NN force AV18 and consistent $\pi$- and $\rho$-like exchange currents we
investigated within the Faddeev framework the process 
$ \overrightarrow{^3{\rm He}} (e,e' {\vec p})d$ 
(or $ \overrightarrow{^3{\rm He}} (e,e' {\vec d})p$).
The aim was to have access to the spin-dependent momentum distribution 
of polarized ${\vec p} {\vec d}$ 
clusters in polarized $^3$He. That distribution would provide interesting 
insight into the $^3$He wave function. We restricted ourselves 
to a nonrelativistic regime, where the 3N c.m. energy
of the final state should stay below the pion threshold. In that kinematical regime we
explored the longitudinal and transverse response functions $W_L$ and $W_T$,
as well as $W_{T'}$,
as a function of
the final deuteron and the allowed photon momenta. All the spins and momenta are
chosen parallel or antiparallel to the photon momentum. While in PWIA $W_L$ and $W_T$ ($W_{T'}$)
up to known factors yield directly the searched for spin-dependent momentum distribution,
FSI and MEC's preclude in most cases the direct access to that distribution. 
The response functions $W_L$ and $W_T$ multiplied by appropriate factors have 
been mapped out in a wide kinematical range and this theoretical outcome
should be checked experimentally. It presents the present day state-of-the-art
insight into the  dominant photon absorption process and the few-nucleon 
dynamics.
It is only at small deuteron momenta 
$p_d  \leq 2 {\rm fm}^{-1} $
that the searched for momentum distribution can be accessed 
within the constrained kinematics we have chosen.

Right now we have no reliable estimate for the amount of relativistic corrections nor
insight into the stability of our results under exchange of nuclear forces and consistent
MEC's. Clearly work in that respect should be envisaged.

\acknowledgements
We are indebted to Dr. H. Gao and Dr. D. Dutta, who inspired
us to perform this study.
We would also like to thank Dr. M.O. Distler for an interesting discussion
about experiments with polarized electrons.
This work was supported by
the Deutsche Forschungsgemeinschaft (J.G.),
the Polish Committee for Scientific Research
and by NFS grant \# PHY0070858.
The numerical calculations have been performed
on the Cray T90 of the NIC in J\"ulich, Germany.


\begin{table}[hbt]
\begin{center}
\begin{tabular}{rrrrrrr} 
\hline
 $\theta_e$ &  $p_N$ & $p_d$ & $\omega$ &  $Q$    & $E_{3N}^{c.m.}$ &  \\
  (deg)       &  (MeV/c)  & (MeV/c)  &  (MeV)     &   (MeV/c) &  (MeV)     &   \\
\hline
     14.45 &      310 &      10 &    56.67 &     300 &    35.22 &  $C_1$ \\ 
     19.43 &      410 &      10 &    95.01 &     400 &    61.14 &  $C_1$ \\ 
     24.56 &      510 &      10 &   144.00 &     500 &    94.15 &  $C_1$ \\ 
     29.91 &      610 &      10 &   203.63 &     600 &   134.26 &  $C_1$ \\ 
     35.58 &      710 &      10 &   273.92 &     700 &   181.47 &  $C_1$ \\ 
           &          &         &          &         &          &        \\
     14.21 &      400 &     100 &    93.33 &     300 &    71.89 &  $C_1$ \\ 
     19.11 &      500 &     100 &   141.26 &     400 &   107.38 &  $C_1$ \\ 
     24.15 &      600 &     100 &   199.83 &     500 &   149.98 &  $C_1$ \\ 
     29.39 &      700 &     100 &   269.05 &     600 &   199.68 &  $C_1$ \\ 
           &          &         &          &         &          &        \\
     19.41 &      200 &     200 &    37.44 &     400 &     3.56 &  $C_2$ \\ 
     24.52 &      300 &     200 &    64.06 &     500 &    14.21 &  $C_2$ \\ 
     29.85 &      400 &     200 &   101.33 &     600 &    31.96 &  $C_2$ \\ 
     35.46 &      500 &     200 &   149.25 &     700 &    56.81 &  $C_2$ \\ 
     41.46 &      600 &     200 &   207.82 &     800 &    88.76 &  $C_2$ \\ 
           &          &         &          &         &          &        \\
     16.93 &       50 &     300 &    30.80 &     350 &     3.58 &  $C_2$ \\ 
     27.05 &      250 &     300 &    62.74 &     550 &     3.58 &  $C_2$ \\ 
     29.70 &      300 &     300 &    77.39 &     600 &     8.02 &  $C_2$ \\ 
     35.22 &      400 &     300 &   114.66 &     700 &    22.21 &  $C_2$ \\ 
     41.10 &      500 &     300 &   162.58 &     800 &    43.51 &  $C_2$ \\ 
           &          &         &          &         &          &        \\
     21.94 &       50 &     400 &    49.45 &     450 &     8.04 &  $C_2$ \\ 
     35.06 &      300 &     400 &    96.04 &     700 &     3.60 &  $C_2$ \\ 
     40.80 &      400 &     400 &   133.32 &     800 &    14.25 &  $C_2$ \\ 
           &          &         &          &         &          &        \\
     14.21 &      200 &     500 &    93.41 &     300 &    71.96 &  $C_3$ \\ 
     19.47 &      100 &     500 &    77.44 &     400 &    43.56 &  $C_3$ \\ 
     24.05 &       10 &     500 &    72.17 &     490 &    24.08 &  $C_3$ \\ 
           &          &         &          &         &          &        \\
     13.31 &      300 &     600 &   149.35 &     300 &   127.90 &  $C_3$ \\ 
     19.28 &      200 &     600 &   122.73 &     400 &    88.86 &  $C_3$ \\ 
     24.62 &      100 &     600 &   106.76 &     500 &    56.91 &  $C_3$ \\ 
     29.32 &       10 &     600 &   101.49 &     590 &    34.23 &  $C_3$ \\ 
\hline
\end{tabular}
\end{center}
\caption[]
{
Electron kinematics together with different kinematical quantities used 
to extract 
the spin dependent momentum distributions
of proton-deuteron clusters in $^3$He.
}
\label{TAB1}
\end{table}


\begin{figure}[h!]
\leftline{\mbox{\epsfxsize=140mm \epsffile{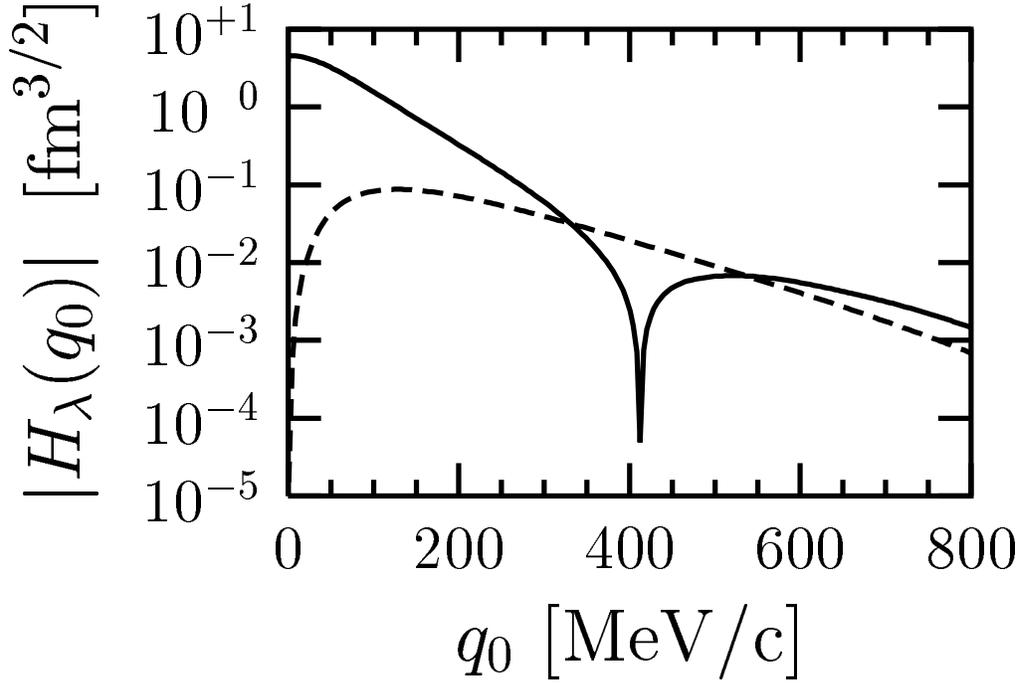}}}
\caption[ ]
{
Absolute value of $ H_\lambda (q_0) $ defined in Eq.~(5) 
for $\lambda=0$ (solid) and $\lambda=2$ (dashed).
Note $ H_0 (q_0) < 0 $ for $ q_0 > $ 400 MeV/c, while 
$ H_2 (q_0) $ remains always positive for the shown 
$q_0$-values.
}
\label{fig1}
\end{figure}

\begin{figure}[h!]
\leftline{\mbox{\epsfxsize=140mm \epsffile{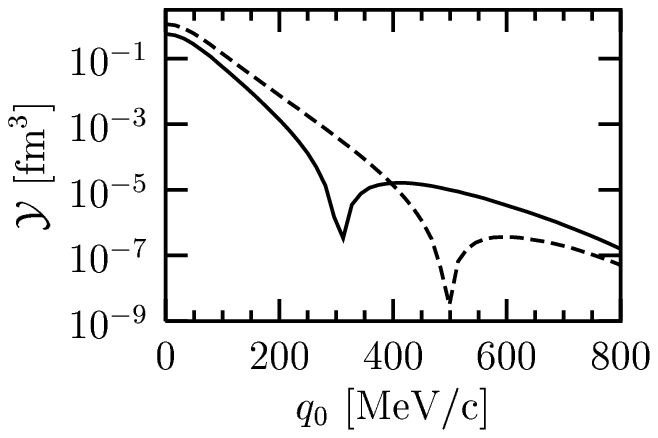}}}
\caption[ ]
{
Spin-dependent momentum distributions 
${\cal Y} ( M= \frac12, M_d= 0, m= \frac12 ; | {\vec q}_0 | {\hat z} )$ 
(solid) 
and
${\cal Y} ( M= \frac12, M_d= 1, m= -\frac12 ; | {\vec q}_0 | {\hat z} )$
(dashed) 
for ${\vec p} {\vec d}$ clusters in $^3$He.
}
\label{fig2}
\end{figure}

\begin{figure}[h!]
\leftline{\mbox{\epsfxsize=140mm \epsffile{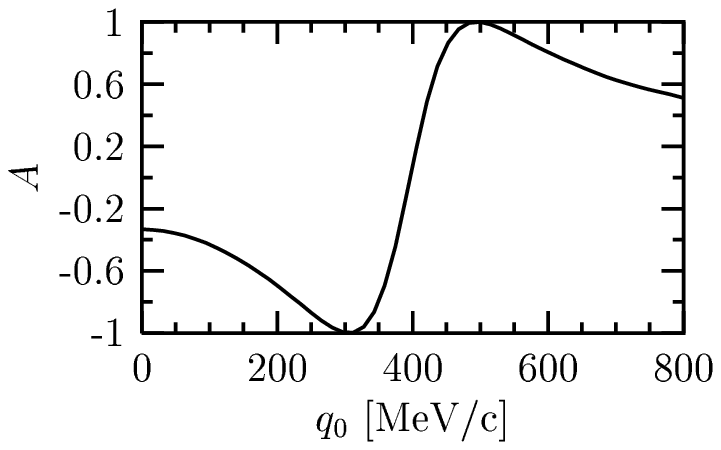}}}
\caption[ ]
{
The asymmetry 
$ A = 
\left( {\cal Y} ( m= \frac12 )  - {\cal Y} ( m= -\frac12 ) \right) /
\left( {\cal Y} ( m= \frac12 )  + {\cal Y} ( m= -\frac12 ) \right) $
}
\label{fig3}
\end{figure}

\begin{figure}[h!]
\leftline{\mbox{\epsfxsize=140mm \epsffile{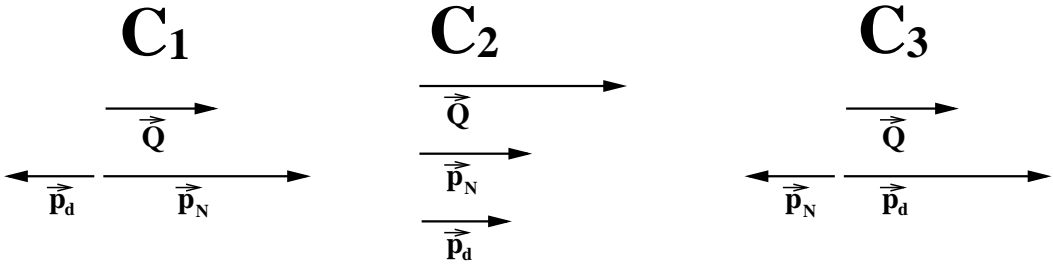}}}
\caption[ ]
{
Three momenta arrangements $C_1$, $C_2$ and $C_3$ for parallel kinematics.
See Table~I.
}
\label{fig4}
\end{figure}

\begin{figure}[h!]
\leftline{\mbox{\epsfxsize=160mm \epsffile{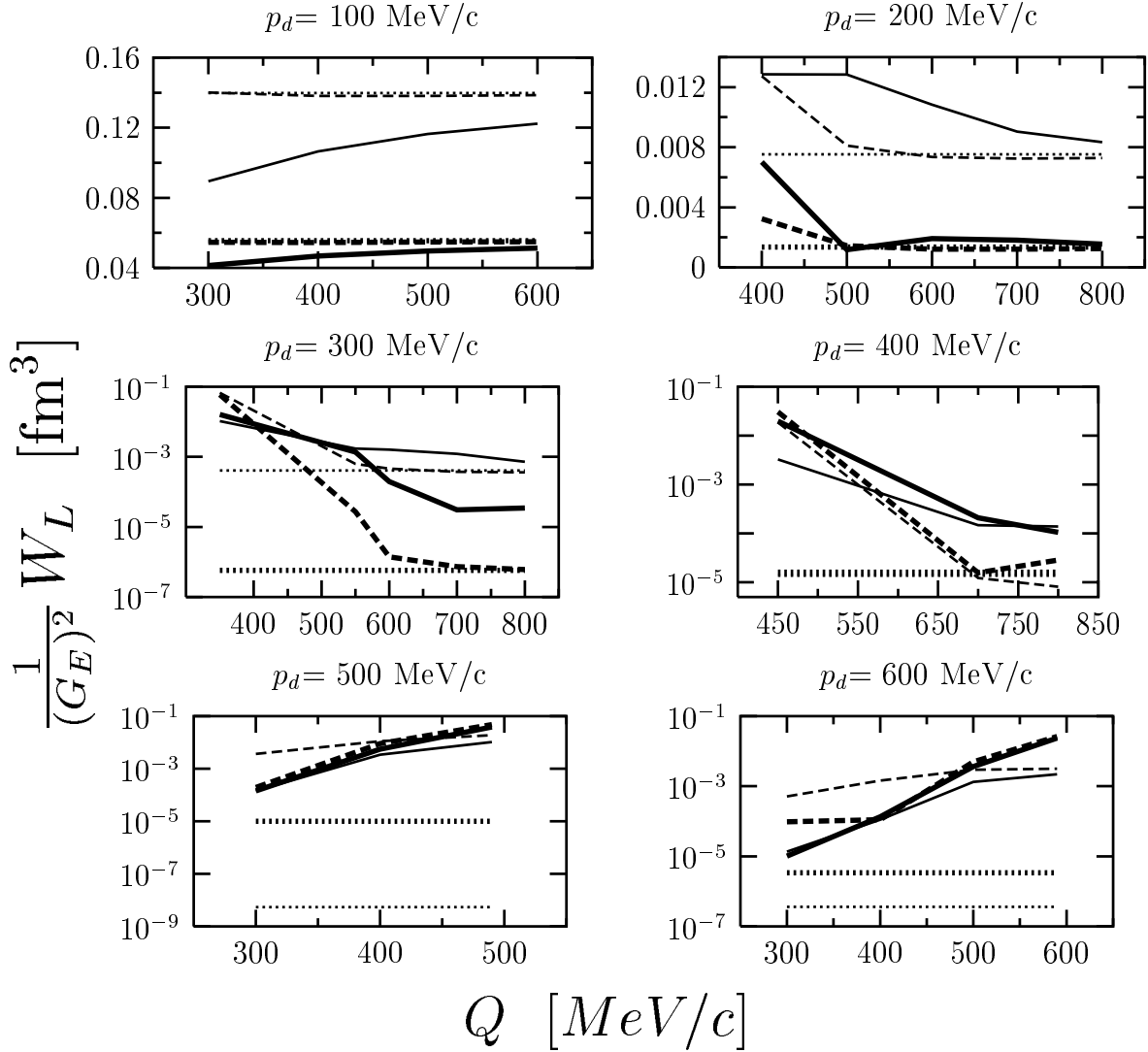}}}
\caption[ ]
{
$\frac1{ (G_E)^2 } \, W_L$ as a function of the three-momentum 
transfer $Q$ 
for different $p_d$-values. 
The curves correspond to PWIA (dotted), PWIAS (dashed)
and Full (solid) results. The thick curves are for the 
$M = \frac12, M_d = 0, m = \frac12$ case, the thin lines for 
the $M = \frac12, M_d = 1, m = -\frac12$ combination of the spin magnetic 
quantum numbers. In case of $p_d$= 400 MeV/c the two PWIA results overlap.
}
\label{fig5}
\end{figure}

\begin{figure}[h!]
\leftline{\mbox{\epsfxsize=160mm \epsffile{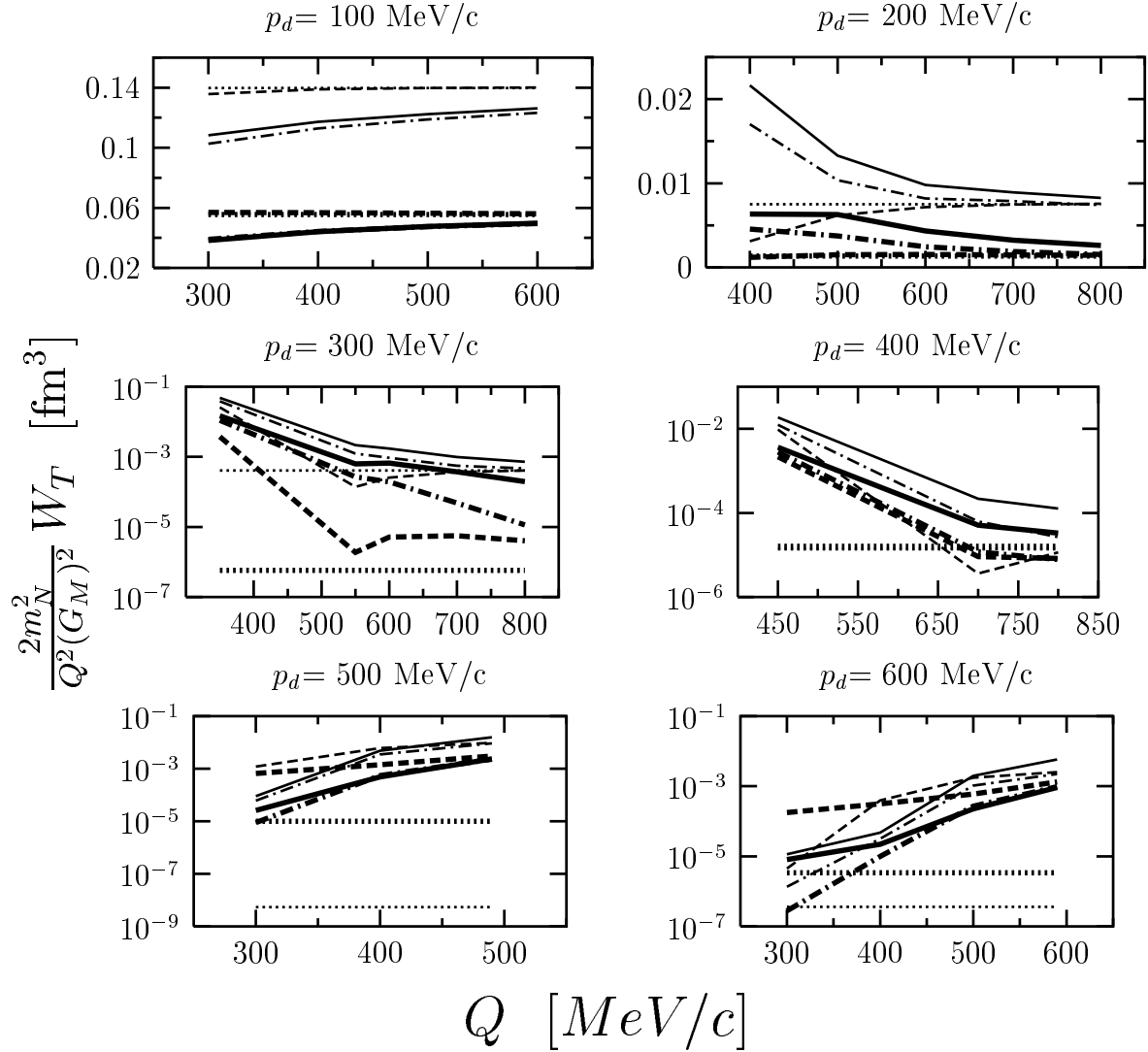}}}
\caption[ ]
{
$\frac{2 m_N^2}{Q^2 (G_M)^2 } \, W_T$  as a function of the three-momentum
transfer $Q$
for different $p_d$-values.
The curves correspond to PWIA (dotted), PWIAS (dashed),
Full without MEC (dash-dotted) and Full including MEC (solid) results. 
The thick curves are for the
$M = \frac12, M_d = 0, m = -\frac12$ case, the thin lines for
the $M = \frac12, M_d = 1, m = \frac12$ combination of the spin magnetic
quantum numbers. In case of $p_d$= 400 MeV/c the two PWIA results overlap.
}
\label{fig6}
\end{figure}

\begin{figure}[h!]
\leftline{\mbox{\epsfxsize=140mm \epsffile{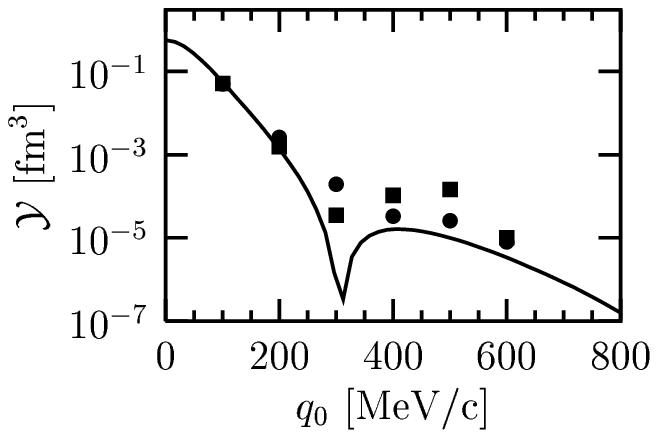}}}
\caption[ ]
{
${\cal Y} ( M = \frac12, M_d = 0, m = \frac12 ; q_0) $ (solid curve)
as a function of the relative proton-deuteron momentum $q_0$
together with the values of closest approach (see text)
from $W_L$ (squares) and from $W_T$ (circles).
}
\label{fig7}
\end{figure}

\begin{figure}[h!]
\leftline{\mbox{\epsfxsize=140mm \epsffile{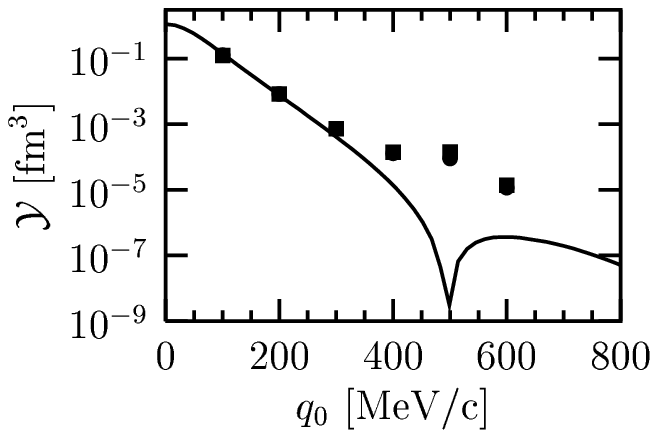}}}
\caption[ ]
{
${\cal Y} ( M = \frac12, M_d = 1, m = -\frac12 ; q_0) $ (solid curve)
as a function of the relative proton-deuteron momentum $q_0$
together with the values of closest approach (see text)
from $W_L$ (squares) and from $W_T$ (circles).
}
\label{fig8}
\end{figure}

\begin{figure}[h!]
\leftline{\mbox{\epsfxsize=140mm \epsffile{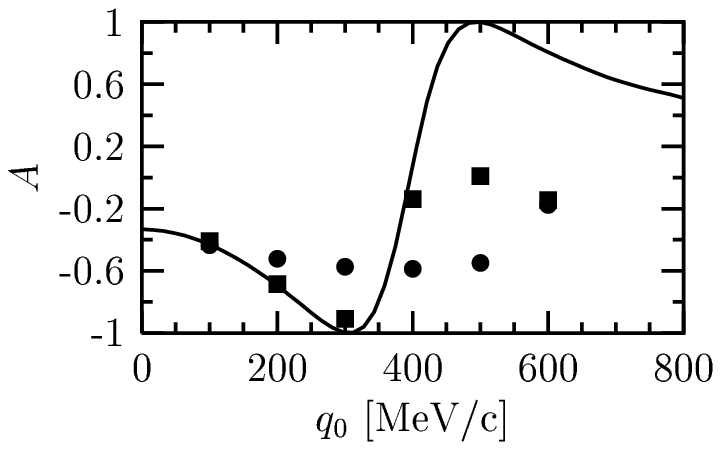}}}
\caption[ ]
{
The asymmetry
$ A =
\left( {\cal Y} ( m= \frac12 )  - {\cal Y} ( m= -\frac12 ) \right) /
\left( {\cal Y} ( m= \frac12 )  + {\cal Y} ( m= -\frac12 ) \right) $
as a function of the relative proton-deuteron momentum $q_0$
together with the values of closest approach (see text)
from $W_L$ (squares) and from $W_T$ (circles).
}
\label{fig9}
\end{figure}

\end{document}